# Generalized linearization in nonlinear modeling of data

W. Chen

Permanent mail address: P. O. Box 2-19-201, Jiangshu University of Science & Technology, Zhenjiang City, Jiangsu Province 212013, P. R. China

Present mail address (as a JSPS Postdoctoral Research Fellow): Apt.4, West 1st floor, Himawari-so, 316-2, Wakasato-kitaichi, Nagano-city, Nagano-ken, 380-0926, JAPAN

E-mail: chenw@homer.shinshu-u.ac.jp

Permanent email: chenwwhy@hotmail.com

## Abstract

The principal innovative idea in this note is to transform the original complex nonlinear modeling problem into a combination of linear problem and very simple nonlinear problems. The key step is the generalized linearization of nonlinear terms. This paper only presents the introductory strategy of this methodology. The practical numerical experiments will be provided subsequently.

## 1. Introduction

The present author [1] presented a generalized linearization strategy in which the nonlinear problems are considered an ill-posed linear system. Consequently, all nonlinear algebraic terms are considered linearly independent variables. A n-dimension nonlinear system can be expanded at most as a linear system of n(n+1)/2 dimension space. In this paper, we extend this idea to nonlinear modeling of data. It is shown that if the interested system only includes polynomial nonlinear terms, the present methodology can apply the linear least square method to handle nonlinear modeling of data.

## 2. Generalized linearization of nonlinear variables

Some linearization procedures such as the Newton-Raphson method are often used in practice to solve nonlinear modeling of data. However, if the nonlinear terms of system equations are considered independent system variables, we can state a n-dimension nonlinear system at most as a n(n+1)/2 dimension linear one.

To better clarify our idea, consider the following example without the loss of generality

$$f(x_1, x_2) = a_1^2 x_1^2 + a_1 a_2 x_1 x_2 + 3a_1 x_1 + 2a_2 x_2 + 2, \tag{1}$$

where $a_1$ and $a_2$ are the desired system parameters. It is noted that there exist two nonlinear terms $a_1^2$ and $a_1 a_2$ in the above equation (1). let

$$b_1 = a_1^2, \tag{2a}$$

$$b_2 = a_1 a_2, \tag{2b}$$

Eq. (1) can be restated as

$$f(x_1, x_2) = b_1 x_1^2 + b_2 x_1 x_2 + 3a_1 x_1 + 2a_2 x_2 + 2, \tag{3}$$

Eq. (3) is a linear parameter identification problem. We can use the normal linear numerical fitting technique such as linear least square method to get desirable parameters.

## 3. Remarks

The nonlinear data modeling has wide applications in the neuro network, data process, and singal process. The presented scheme greatly simplifies the solution process. Some numerical examples will be subsequently provided to demonstrate the effectiveness of this strategy.

**Reference**

1. W. Chen, Generalized linearization of nonlinear algebraic equations: an innovative approach, published in http://xxx.lanl.gov/abs/math.NA/9905042